\documentstyle[pre,twocolumn,aps,epsf]{revtex}

\begin{document}
\title{Synchronization in a ring of pulsating oscillators with bidirectional
couplings}
\author{Albert D\'{\i}az-Guilera\thanks{albert@ffn.ub.es},
Conrad J. Perez-Vicente\thanks{conrad@ffn.ub.es}}
\address{
Departament de F\'{\i}sica Fonamental, 
Universitat de Barcelona, 
Diagonal 647, E-08028 Barcelona, Spain 
}
\author{Alex Arenas\thanks{aarenas@etse.urv.es}}
\address{
Departament d'Enginyeria Inform\`{a}tica, Universitat
Rovira i Virgili, Carretera Salou s/n, E-43006 Tarragona, Spain 
}
\date{\today}

\maketitle

\begin{abstract}
We study the dynamical behavior of an ensemble of oscillators interacting
through short range bidirectional pulses. The geometry is 1D with periodic
boundary conditions. Our interest is twofold. To explore the conditions
required to reach fully synchronization and to invewstigate the time
needed to get such state. We present both theoretical and numerical results.
\end{abstract}




The analysis of the dynamic properties of populations of pulse-coupled
oscillators are the starting point of many studies devoted to understand
some phenomena such as synchronization, phase locking or the emergence of
spatio-temporal patterns which appear so frequently when analyzing the
behavior of heart pacemaker cells, integrate and fire neurons, and other
systems made of excitable units [Peskin, 1984 ], [Mirollo \& Strogatz, 
1990], [Kuramoto, 1991], [Abbott \& Van Vreeswijk, 1993], [Treves, 1993].

Mean-field models or populations of just a few oscillators are the typical
subject that has been considered in the scientific literature. In these
simplified systems it is possible to investigate analytically the main
mechanisms leading to the formation of assemblies of synchronized elements
as well as other spatio-temporal structures. However, such restrictions do
not allow to consider the effect of certain variables whose effect can be
crucial for realistic systems. The specific topology or geometry of the
system, as well as the precise connectivity between units are some typical
examples which usually induce important changes in the collective behavior
of these models. Unfortunately, a rigorous mathematical description of them
is still missing. The majority of studies rely on simulations showing the
outstanding richness that a low dimensional system of pulse-coupled
oscillators may display. Some examples are self-organized criticality,
chaos, quasiperiodicity, etc. [Perez et al., 1996]. In other cases, it is
proven the stability of some observed behaviors [Goldsztein \& Strogatz,
1995], [Diaz Guilera et al., 1997] but not the mechanisms leading to them.

A first step forward has been given very recently by [Diaz Guilera et al.,
1998], hereafter called DPA. Assuming a system defined on a ring, they
developed a mathematical formalism powerful enough to get analytic
information not only about the mechanisms which are responsible for
synchronization and formation of spatio-temporal structures, but also, as a
complement, to proof under which conditions they are stable solutions of the
dynamical equations. They consider one-directional interactions which allow
to simplify the analysis. The study of a more general situation is
desirable.

The aim of this letter is to show that such formalism is able to handle more
difficult situations. In particular, we consider here a population of
pulse-coupled oscillators with bidirectional couplings. This fact is
important because the backwards effect of the coupling might break the
coherent activity of an ensemble of oscillators previously synchronized.
Small changes in local aspects of the coupling may lead to important
cooperative effects.

Let us start the discussion by introducing the model and the notation used
throughout the paper. We have considered a population of $(N+1)$ pacemakers
distributed on a ring. This geometry is interesting to analyze certain
problems related to cardiac activity. For instance, to study some types of
cardiac arrhythmia characterized by an abnormally rapid heartbeat whose
period is set by the time that an excitation takes to travel a circuit. This
observation can be explained by modeling appropriately the circulation of a
wave of excitation in a one-dimensional ring [Ito \& Glass, 1992]. The
study of the sinus rhythm has been also studied in systems with similar
geometry and bidirectional couplings [Ikeda, 1982]. Other systems whose
dynamical evolution is restricted to a limit cycle and therefore that can be
described in terms of only one degree of freedom could also be tackle with
the same tools.

The usual description of the state of one unit is performed in terms of one
physical variable that, in general, it is voltage-like quantity. However,
after a straightforward transformation [Corral et al., 1995] it is always
possible to write the evolution of each oscillator in terms of a phase
variable $\phi \in [0,1]$ which evolves linearly in time. In addition, the
effect of pulsating-interaction between oscillators can be written down
through the so called phase response curve (PRC) which measures the
effective change in $\phi$ due to the firing process. In general, all the
nonlinearities of the problem are included in this function. In this letter,
we are interested in monotonic functions which are closely related to the
convex character of the evolution of the voltage like variable in formal
pacemakers (perfect integrators). Within this space of functions we have
chosen the linear case because is allows to get analytical results without
losing any of the features which characterize the synchronization process
between units. Let us clarify this point. When a given unit reaches a
threshold value $\phi _{th}=1$, it fires and changes the state of its
neighbors according to 
\[
\phi _{i}\geq 1\Rightarrow \left\{ 
\begin{array}{l}
\phi _{i}\rightarrow 0 \\ 
\phi _{nn}\rightarrow \phi _{nn}+\varepsilon \phi _{nn}\equiv \mu \phi _{nn}
\end{array}
\right. 
\]
\begin{equation}
\hspace{5em}\forall i=0,\ldots ,N
\end{equation}
where $nn$ denotes the nearest neighbors, $\varepsilon $ is the strength of
the coupling and where $N+1\equiv 0$ due to the boundary conditions.
According to the aforementioned definition $\mu \phi$ is the PRC.

One of the key points of the mathematical formalism relies on a suitable
transformation which allows to trace the phases of the oscillators after
each firing and construct return maps of the complete cycle. For details see
[Corral et al., 1995]. The transformation which intrinsically includes
translations and rotations always keeps information about the element of the
population which will fire immediately. This is a very appropriate way to
know details about the spatio-temporal structure which is dynamically
forming at every time step.

From our point of view the most direct way to understand the mechanism
underlying synchronization or any other time-dependent phenomenon is to
start analyzing the simplest situation with physical interest. For a
population with bidirectional couplings such case is to consider only four
oscillators. In DPA and for the one-directional case the same scheme was
applied. However, they started with even a more simpler situation, since
only three elements were matter of attention. For bidirectional interactions
such case is trivial due to the symmetry of the problem since one firing
affect all the neighbors and from an effective point of view the problem is
mean-field which has been already solved in [Mirollo \& Strogatz, 1990].

For the four pacemakers system there are six possible sequences of firings
such that one member of the population fires once and only once in each
cycle. If we assume that the oscillator which fires is always labeled as
unit $0$ and the rest of elements are ordered from this unit clockwise then
these sequences are the following:

\begin{itemize}
\item  A: 0,1,2,3

\item  B: 0,1,3,2

\item  C: 0,2,1,3

\item  D: 0,2,3,1

\item  E: 0,3,1,2

\item  F: 0,3,2,1
\end{itemize}

where the sequence has to be understood as the order in which oscillators
reach the threshold. The dynamical process of firings can be described in
terms of a set of matrices which take into account the 'jump' (distance)
between those oscillators which fire consecutively [Diaz Guilera et al,
1998]. It is straightforward to check that they are

\begin{equation}
M_1=\left( 
\begin{array}{ccc}
-\mu & 1 & 0 \\ 
-\mu & 0 & \mu \\ 
-\mu & 0 & 0
\end{array}
\right)
\end{equation}

\begin{equation}
M_2=\left( 
\begin{array}{ccc}
0 & -1 & \mu \\ 
0 & -1 & 0 \\ 
\mu & -1 & 0
\end{array}
\right)
\end{equation}

\begin{equation}
M_3=\left( 
\begin{array}{ccc}
0 & 0 & -\mu \\ 
\mu & 0 & -\mu \\ 
0 & 1 & -\mu
\end{array}
\right)
\end{equation}
and that a all the cycles related to the different orders are constructed by
combining the three previous matrices as follows

\begin{itemize}
\item  A: 0,1,2,3 $\rightarrow $ $T_{1}\circ T_{1}\circ T_{1}\circ T_{1}$

\item  B: 0,1,3,2 $\rightarrow $ $T_{2}\circ T_{3}\circ T_{2}\circ T_{1}$

\item  C: 0,2,1,3 $\rightarrow $ $T_{1}\circ T_{2}\circ T_{3}\circ T_{2}$

\item  D: 0,2,3,1 $\rightarrow $ $T_{3}\circ T_{2}\circ T_{1}\circ T_{2}$

\item  E: 0,3,1,2 $\rightarrow $ $T_{2}\circ T_{1}\circ T_{2}\circ T_{3}$

\item  F: 0,3,2,1 $\rightarrow $ $T_{3}\circ T_{3}\circ T_{3}\circ T_{3}$
\end{itemize}

where $T_{i}$ is defined as 
\[
{\vec{\phi}}^{\prime}= T_i(\vec{\phi}) \equiv \vec{1}+M_i \vec{\phi}, 
\]
where $\vec{\phi}^{\prime}$ is a vector with $N$ components since the
zero-th component does not play any role in the description.

Before computing the fixed points of the transformation as well as the
stability of the associated eigenvalues let us notice that matrices $M_{1}$, 
$M_{2}$ and $M_{3}$ have exactly the same structure as in the
one-directional coupling case except for one column which is multiplied by $%
\mu $. Therefore many properties of the new matrices can be discussed
directly without the explicit calculation of them. In particular in DPA it
was shown that the resultant matrices had eigenvalues with moduli larger or
smaller than 1 depending on the sign of the coupling, i.e.e $\varepsilon $,
which is rather important because it determines the stability of the fixed
points and as a consequence ensures that for excitatory couplings the
oscillators will synchronize their activity while for the inhibitory case
complex spatio-temporal structures will be formed in the stationary state.
Now, these properties do not change at all. The nature of the eigenvalues
does not change and independently of the particular position of the new
fixed points the conditions for stability are the same as in the
one-directional case. Even more, it is straightforward to show that the
modulus of the eigenvalues that in the aforementioned case were larger than $%
1$ are now larger and the opposite for those which are smaller than one.
This fact suggests that the stability of the typical spatio-temporal
structures found in DPA such as the chessboard configuration are more stable
for the bidirectional case and the opposite in the other side, i.e.e, units
synchronize faster.

The generalization to an arbitrary number of oscillators is a technical
matter that follows the same steps discussed in DPA. For this reason the
mathematical details will not be discussed here again. However, we want to
stress that according to these results the physical scenario defined for a
population with one-directional couplings still holds. The oscillators
synchronize due to a mechanism of dimensional reduction. However, let us
remark an important difference respect to the previous case. In the
bidirectional case one cannot ensure that two units firing at unison in a
given cycle will do it again in the next one, however it can be ensured
through simple algebra that the number of synchronized oscillators cannot
decrease at any time, in other words, if a couple break their mutual
synchrony necessarily one of the elements of this couple will synchronize
with the other neighbor. This fact shows that the term 'dimensional
reduction' is more appropriate than the word 'absorption' used currently in
the literature. On the other hand, in the typical inhibitory patterns the
oscillators tend to be as far as possible from the nearest neighbors.

As a complement of the previous studies we have analyzed the time required
to reach fully synchronization for the bidirectional case. In particular we
have focused our attention in two particular situations. It is well known
that for only two oscillators the maximal time needed to find both moving in
synchrony is when in the initial condition the units are separated each
other by $\phi =0.5$, i.e.e when they are separated as far as possible. We
wanted to check whether such phenomenon is also observed for larger
populations. To do this, we have assumed that the oscillators are
distributed randomly in a interval $[\phi _{min},1]$ where $\phi _{min}$ is
a variable quantity. Figure 1 shows the results for different coupling
strengths $\varepsilon $ averaged over 200 samples. As we can see the
randomization breaks the singular character of $\phi =0.5$ and the time
grows monotonically with the width of the interval.

\begin{figure}[h]
\center{
\epsfxsize=0.9\linewidth
\epsffile{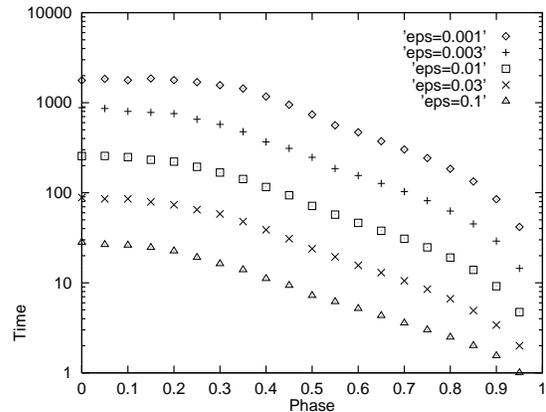}
}
\caption{Time needed to reach fully synchronization versus $\phi_{min}$ for
several values of the coupling. The results are an average over 200 samples.
The results are given for $N=32$.}
\end{figure}

We have also studied for a fixed interval and for a fixed number of elements
how depend the synchronization time with the magnitude of the coupling. In
the mean-field case it was shown [Strogatz et al., 1990] that the functional
dependence goes as $t \approx 1/ \varepsilon$. Figure 2 shows our results
for the bidirectional case. It is interesting to see that the same
relationship between both variables still holds for short range couplings.

\begin{figure}
\center{
\epsfxsize=0.9\linewidth
\epsffile{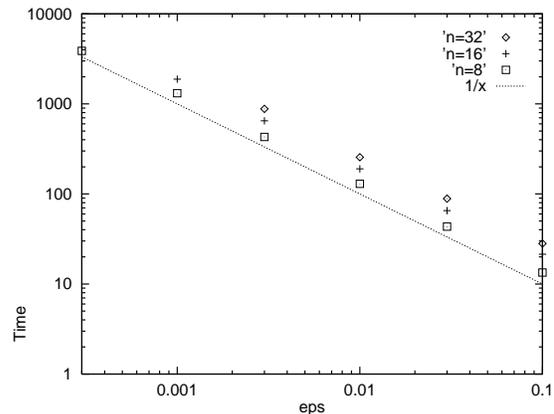}
}
\caption{Time needed to reach fully synchronization versus magnitude of the
coupling for different population sizes. The results are an average over 200
samples. We have also plot the $1/x$ line as a guide for the eye.}
\end{figure}

In conclusion, we have studied under which conditions a population of
pulse-coupled oscillators with bidirectional interactions display either
synchronization or spatio-temporal structures. We have followed the
mathematical formalism developed in DPA for a simpler case, noticing that
the physical mechanisms underlying both phenomena are the same (dimensional
reduction) as in the one-directional situation. We have also studied the
time required to get synchronization observing the same functional
dependence in the coupling found for the mean-field approach.

\acknowledgments
This work has been supported by CICyT of the Spanish Government, grant
\#PB96-0168.

\section*{references}

\noindent L.F. Abbott and C. Van Vreeswijk, [1993] "Asynchronous states in
networks of pulse-coupled oscillators" {\em Phys. Rev. E} {\bf 48},
1483-1490.

\noindent A. Corral, C.J. P\'{e}rez, A. D\'{\i }az-Guilera, and A. Arenas,
[1995] ''Synchronization in a lattice model of pulse-coupled oscillators'', 
{\em Phys. Rev. Lett}. {\bf 75}, 3697-3700.

\noindent A. D\'{\i}az-Guilera, A. Arenas, A. Corral, and C.J.
P\'{e}rez,[1997] "Stability of spatio-temporal structures in a lattice model
of pulse-coupled oscillators", {\em Physica} {\bf 103D}, 419-429.

\noindent A. D\'{\i }az-Guilera, A. Arenas, and C.J. P\'{e}rez, [1998]
''Mechanisms of synchronization and pattern formation in a ring of
coupled-oscillators'', {\em Phys. Rev. E} {\bf 57}, 3820

\noindent G. Goldsztein and S.H. Strogatz, [1995] {\em Int. J. Bifurcation
and Chaos} {\bf 5}, 983

\noindent H. Ito and L. Glass, [1992] Physica {\bf 56D}, 84.

\noindent N. Ikeda, [1982] "Model of bidirectional interaction between
myocardial pacemakers based on the phase response curve" {\em Biol. Cybern.} 
{\bf 43}, 157-167.

\noindent Y. Kuramoto, [1991] " Collective synchronization of pulse-coupled
oscillators and excitable units", {\em Physica} {\bf 50D}, 15-30.

\noindent R. Mirollo and S.H. Strogatz, [1990] "Synchronization of
pulse-coupled biological oscillators" {\em SIAM J. Appl. Math.} {\bf 50},
1645-1662.

\noindent C.J. P\'{e}rez, A. Corral, A. D\'{\i}az-Guilera, K. Christensen,
and A. Arenas, [1996], "Self-organized criticality and synchronization in
lattice models of coupled dynamical systems" {\em Int. J. Mod. Phys} {\bf 10}%
, 1111-1151.

\noindent C.S. Peskin, [1984] {\em Mathematical Aspects of Heart Physiology}%
, Courant Institute of Mathematical Sciences, New York University, (New
York) 268.

\noindent M. Sousa Vieira, A.J. Lichtenberg, and M.A. Lieberman, [1994] Int.
J. Bifurcation and Chaos {\bf 4}, 1563.

\noindent A. Treves, [1993] "Mean-field Analysis of neuronal spike dynamics" 
{\em Network} {\bf 4}, 259-284.

\end{document}